\documentclass[preprint,12pt,number]{elsarticle}

\usepackage{lineno,hyperref}
\usepackage{amssymb}
\usepackage{amsmath}
\usepackage{bm}
\usepackage{graphicx}
\usepackage{booktabs}
\usepackage{subcaption}
\usepackage[section]{placeins}
\usepackage{algorithm}
\usepackage{algorithmic}

\usepackage{tikz}
\usetikzlibrary{arrows.meta,positioning,calc}

\modulolinenumbers[5]
\relax
\begin{document}

\begin{frontmatter}



\title{Macroscopic transport patterns of UAV traffic in 3D anisotropic wind fields: A constraint-preserving hybrid PINN–FVM approach} 


\author[1]{Hanbing Liang} 
\author[1]{Fujun Liu\corref{cor1}}
\ead{fjliu@cust.edu.cn}
\affiliation[1]{organization={Nanophotonics and Biophotonics Key Laboratory of Jilin Province, School of Physics, Changchun
University of Science and Technology},
            addressline={}, 
            city={Changchun},
            postcode={130022}, 
            state={},
            country={P.R. China}}

\begin{abstract}
Macroscopic unmanned aerial vehicle (UAV) traffic organization in three-dimensional airspace faces significant challenges from static wind fields and complex obstacles. A critical difficulty lies in simultaneously capturing the strong anisotropy induced by wind while strictly preserving transport consistency and boundary semantics, which are often compromised in standard physics-informed learning approaches. To resolve this, we propose a constraint-preserving hybrid solver that integrates a physics-informed neural network for the anisotropic Eikonal value problem with a conservative finite-volume method for steady density transport. These components are coupled through an outer Picard iteration with under-relaxation, where the target condition is hard-encoded and strictly conservative no-flux boundaries are enforced during the transport step. We evaluate the framework on reproducible homing and point-to-point scenarios, effectively capturing value slices, induced-motion patterns, and steady density structures such as bands and bottlenecks. Ultimately, our perspective emphasizes the value of a reproducible computational framework supported by transparent empirical diagnostics to enable the traceable assessment of macroscopic traffic phenomena.
\end{abstract}

\begin{graphicalabstract}
\includegraphics[width=\linewidth]{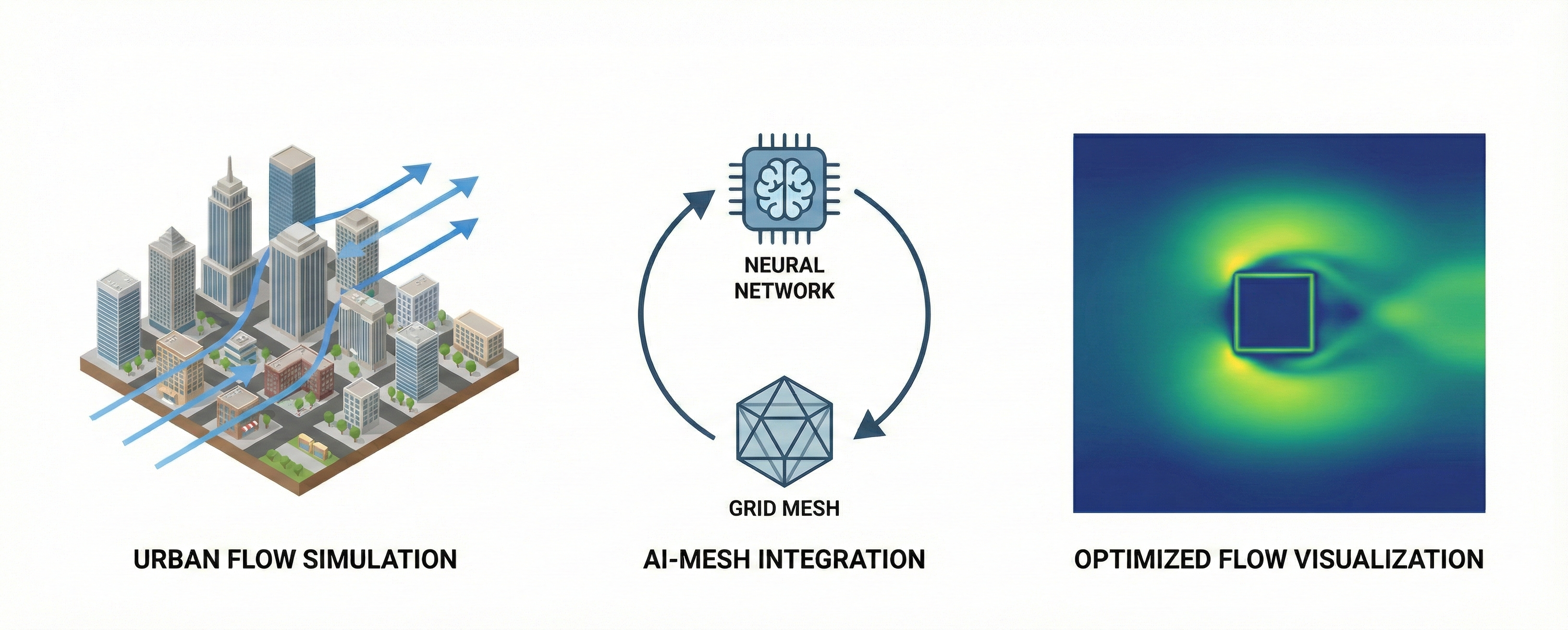}
\end{graphicalabstract}

\begin{highlights}
\item A hybrid PINN-FVM framework models 3D macroscopic UAV transport in wind. 

\item Anisotropic Eikonal equation captures wind-induced symmetry breaking in travel time.

\item Conservative flux discretization ensures physical consistency and strict mass balance. 

\item Steady-state density bands and bottlenecks emerge from coupled field interactions.

\end{highlights}


\end{frontmatter}



\section{Introduction}

Urban air mobility and large-scale unmanned aerial vehicle (UAV) operations motivate the need for traffic-level planning and analysis that extends beyond single-vehicle path planning \cite{thipphavong2018uam,kopardekar2016utm}. As flight tasks become frequent and numerous vehicles operate simultaneously, it becomes practical to summarize collective behavior using macroscopic fields defined over the airspace. In this work, we focus on generating three specific outputs: a value field that encodes minimum-time reachability to a target, an induced macroscopic motion field that provides an interpretable preferred direction of travel, and a steady density distribution that highlights areas of traffic concentration. From an engineering perspective, these fields provide a compact interface between planning and evaluation that is directly actionable for corridor design, bottleneck identification, and comparative analysis under environmental disturbances. Instead of simulating large numbers of individual agents with detailed dynamics, we can study how wind and obstacles reshape travel-time geometry and how these changes translate into congestion-like density patterns, which leads us to emphasize reproducible evaluation with explicit diagnostics in three-dimensional environments.

Addressing this problem requires overcoming two interacting challenges. First, wind fields introduce strong anisotropy into travel-time costs, leading to anisotropic Eikonal and Hamilton-Jacobi-Bellman type equations whose solutions can vary sharply across space \cite{sethian2003ordered}. Second, obstacles impose geometric constraints that require consistent boundary semantics, meaning obstacle surfaces should act as no-flux boundaries for density transport while the target serves as an absorbing region. Classical grid-based partial differential equation solvers can provide strong numerical control but may be costly to implement and tune for complex geometries and heterogeneous coefficients. In contrast, physics-informed neural networks (PINNs) offer flexibility through automatic differentiation and geometry-aware sampling \cite{raissi2019pinn}, making them attractive for the value-function subproblem. However, end-to-end PINN formulations that also learn the transport-dominated density equation often struggle to enforce mass conservation and flux boundary conditions in a quantitatively stable way \cite{mishra2020pinnerror}, especially when sharp gradients or narrow passages are present.

Given a bounded 3D domain with obstacles, an absorbing target region, and a static wind field, we compute a minimum-time value function \(\phi(\bm{x})\), the induced macroscopic velocity \(\bm{u}(\bm{x})\), and a steady density field \(\rho(\bm{x})\). We focus on steady-state behavior under static wind and scenario-defined sources, placing our primary emphasis on a reproducible computational framework and empirical diagnostics rather than theoretical well-posedness or convergence guarantees. To ensure the work remains traceable, all numerical values and hyperparameters referenced in the paper are sourced directly from configuration files. Consequently, each experiment produces run artifacts, including configuration snapshots, metrics logs, metadata, and plots, which support post hoc inspection and verification.

\subsection{Contributions}
The primary contribution of this work is a constraint-preserving hybrid computational route designed to address the distinct numerical challenges inherent in the coupled continuum medium. We employ a physics-informed neural network to solve the value-function subproblem, which effectively handles the complexities of three-dimensional anisotropic settings where standard grid-based methods often struggle. Simultaneously, we utilize a conservative finite-volume method to solve the steady density-transport equation. This specific combination allows for explicit and rigorous control over mass conservation and flux boundary semantics, ensuring that the resulting density fields remain physically meaningful even in the presence of complex obstacles.

To integrate these distinct solvers, we develop a practical coupling strategy for the value-density fixed point that relies on an outer Picard iteration scheme. We implement under-relaxation to mitigate numerical oscillations that typically arise from the strong coupling between congestion and speed. Furthermore, the framework includes a comprehensive set of recorded diagnostics, such as Eikonal residual statistics and density solver residuals. These metrics provide a transparent record of solver behavior, allowing users to perform empirical stability checks and verify that the continuum medium has reached a consistent steady state without relying solely on theoretical assumptions.

Finally, we validate the framework through a systematic and reproducible evaluation suite that covers diverse homing and point-to-point scenarios across multiple wind types and obstacle configurations. Unlike studies that rely solely on qualitative snapshots, our approach derives all reported results directly from traceable run artifacts, including precise configuration files and detailed metric logs. This emphasis on reproducibility ensures that the computed value slices, induced motion patterns, and steady density structures can be rigorously inspected and verified by the broader research community.

\section{Related work}

This research sits at the intersection of three distinct yet complementary areas: continuum traffic-flow modeling with boundary-flux semantics, value-function formulations for routing under anisotropic costs, and physics-informed learning for partial differential equations (PDEs). Rather than proposing a new PDE model or providing theoretical guarantees, our primary contribution is a reproducible, constraint-aware computational framework supported by systematic empirical evaluation. Our design is grounded in the practical numerical observation that the coupled problem combines two diverging characteristics. The value-function PDE benefits significantly from smooth approximations and flexible sampling, whereas the transport equation relies heavily on conservative flux bookkeeping and strict boundary semantics. This distinction motivates a solver architecture that utilizes specialized numerical tools for each subproblem.

\subsection{Continuum flow models and conservation laws}
Macroscopic flow models summarize collective motion using a density field governed by a continuity equation \cite{lighthill1955kinematic,wang2022complexity}. These models serve as widely used, interpretable abstractions for large populations, yielding outputs such as steady density bands, bottlenecks, and splitting patterns that offer a traffic-oriented perspective on airspace organization. In transport-dominated regimes, reliably enforcing mass conservation is more than a numerical detail, as violations can directly translate into misleading density accumulation or artificial depletion. Similarly, boundary flux semantics are critical in obstacle-rich environments, where enforcing no-penetration at obstacles and absorption at targets encodes the intended physical meaning of the scenario.

\subsection{Value functions under anisotropic costs}
Minimum-time routing in the presence of external fields naturally leads to anisotropic Eikonal or Hamilton-Jacobi-Bellman (HJB) type equations \cite{sethian2003ordered,falcone2013semilagrangian}. Wind introduces direction-dependent effective costs that reshape the geometry of optimal trajectories, effectively changing which passages attract traffic and where congestion patterns form. In three-dimensional domains, the resulting value field often varies sharply around obstacles and within strong-wind regions, creating a need for solvers that can represent anisotropy without requiring bespoke discretization for every new wind model.

\subsection{Coupled value--density fixed points}
The interaction between value fields and traffic density creates a fixed-point loop where density affects effective speed, speed modifies the value function, the value induces a macroscopic velocity, and this velocity updates the density. Even when the individual subproblems are robust in isolation, the coupled continuum medium remains sensitive to numerical choices. Consequently, practical solvers often employ alternating Picard iterations with relaxation, alongside empirical diagnostics and stopping criteria to ensure stability. In the Mean Field Game (MFG) literature, stationary models similarly couple an HJB equation with a continuity equation via congestion-dependent dynamics \cite{lasry2007mfg,achdou2010mfgnum}. Our system is closely related to this lineage and can be viewed as an engineering-oriented instantiation for macroscopic UAV traffic. However, rather than pursuing game-theoretic equilibrium analysis or theoretical guarantees such as existence and uniqueness, we treat the coupling as a practical computation supported by reproducible diagnostics, a perspective that aligns with other applied transportation settings \cite{she2024truckdrone}.

\subsection{PINNs and limitations for conservation laws}
PINNs provide flexibility for high-dimensional PDEs and complex coefficients, making them particularly attractive for solving value-function problems in 3D using geometry-aware sampling \cite{raissi2019pinn}. Despite these strengths, conservation-law constraints and flux boundary conditions are challenging to satisfy in a stable and quantitative manner when trained end-to-end \cite{mishra2020pinnerror}, especially in the presence of sharp transport features or complex obstacle boundaries. While some research attempts to enhance PINN stability via variational or hp-style formulations \cite{kharazmi2020hpvpinns}, the hybrid route adopted in this paper leverages PINNs where they are strongest, specifically for the anisotropic value PDE, while retaining conservative discretization for density transport to explicitly preserve physical semantics.

\section{Problem formulation}

\subsection{Domain, obstacles, and absorbing target}
We consider a bounded airspace domain \(\Omega\subset\mathbb{R}^3\) partitioned into an obstacle region \(\Omega_{obs}\) and free space \(\Omega_{free}=\Omega\setminus\Omega_{obs}\). Within the free space, we define a small absorbing target region \(\mathcal{D}\subset\Omega_{free}\), whose boundary \(\partial\mathcal{D}\) serves as the interface where traffic exits the system. In our specific experimental scenarios, \(\Omega\) is represented as a rectangular box, and obstacles are modeled as axis-aligned bounding boxes. To ensure reproducibility and separation of concerns, all specific geometric parameters, including domain bounds, target placement, and obstacle lists, are defined externally in configuration files located at \texttt{paperconfig/*.yaml}. This approach allows us to focus the mathematical description on the intended physical semantics and boundary conditions while delegating numerical specifics to the configuration artifacts.

\subsection{Individual dynamics and time-optimal control}
The motion of an individual vehicle is modeled as the superposition of an air-relative control velocity \(\bm{v}_r\) and a static background wind field \(\bm{v}_w\). The governing dynamics and the speed constraint are given by
\begin{equation}
\dot{\bm{x}}(t)=\bm{v}_r(t)+\bm{v}_w(\bm{x}(t)),\qquad \|\bm{v}_r(t)\|\le v_{max}(\rho(\bm{x}(t))).
\end{equation}
The objective is to minimize the arrival time to the target set \(\mathcal{D}\), which corresponds to a running cost of \(\int_0^T 1\,dt\). The minimum remaining time required to reach \(\mathcal{D}\) from a state \(\bm{x}\) is denoted by the value function \(\phi(\bm{x})\).

\subsection{Congestion-dependent speed and strong controllability}
Macroscopic congestion effects are incorporated through a fundamental diagram that links local density to the maximum achievable airspeed \(v_{max}(\rho)\). For our experiments, we employ a smooth variant of the Greenshields model, implemented via a log-sum-exp smooth maximum function:
\begin{equation}
v_{max}(\rho)=\mathrm{SmoothMax}_\beta\Bigl(v_{min},\ v_{max}^0\bigl(1-\rho/\rho_{jam}\bigr)\Bigr),\qquad
\mathrm{SmoothMax}_\beta(a,b)=\tfrac{1}{\beta}\log\bigl(e^{\beta a}+e^{\beta b}\bigr).
\end{equation}
This formulation includes optional clipping to enforce scenario-specific speed ranges. To ensure that the target remains reachable regardless of wind conditions, we adopt a strong controllability assumption. Specifically, we assume that for all locations \(\bm{x}\in\Omega\) and operational densities \(\rho\in[0,\rho_{max}]\), the vehicle possesses sufficient airspeed to overcome the wind field with a safety margin \(\epsilon_c\):
\begin{equation}
v_{max}(\rho)>\|\bm{v}_w(\bm{x})\|+\epsilon_c.
\end{equation}
This modeling assumption defines the valid operating regime, with all associated parameters drawn from the configuration files.

\subsection{Value function and induced velocity}
Based on the dynamics and objectives defined above, the dynamic programming principle leads to a stationary HJB equation:
\begin{equation}
\min_{\|\bm{v}_r\|\le v_{max}(\rho)}\Bigl\{\nabla\phi(\bm{x})\cdot(\bm{v}_r+\bm{v}_w(\bm{x}))+1\Bigr\}=0.
\end{equation}
The time-optimal control strategy saturates the speed constraint and aligns opposite to the value gradient direction. Formally, the optimal air-relative velocity is
\begin{equation}
\bm{v}_r^*(\bm{x})=-v_{max}(\rho(\bm{x}))\,\frac{\nabla\phi(\bm{x})}{\|\nabla\phi(\bm{x})\|_{\epsilon_{reg}}}.
\end{equation}
Substituting this optimal control back into the HJB equation yields the anisotropic Eikonal equation used in our framework:
\begin{equation}
v_{max}(\rho)\,\|\nabla\phi\|_{\epsilon_{reg}}-\bm{v}_w\cdot\nabla\phi=1,\qquad \bm{x}\in\Omega_{free}\setminus\mathcal{D},
\end{equation}
subject to the absorbing boundary condition \(\phi=0\) on \(\partial\mathcal{D}\). To prevent numerical singularities in regions where the gradient \(\|\nabla\phi\|\) approaches zero, we employ a regularized norm \(\|\bm{z}\|_{\epsilon_{reg}}=\sqrt{\|\bm{z}\|^2+\epsilon_{reg}^2}\), where \(\epsilon_{reg}\) is a configurable parameter.

The resulting macroscopic velocity field \(\bm{u}(\bm{x})\) describes the effective ground velocity of the traffic flow:
\begin{equation}
\bm{u}(\bm{x})=\bm{v}_w(\bm{x})-v_{max}(\rho(\bm{x}))\frac{\nabla\phi(\bm{x})}{\|\nabla\phi(\bm{x})\|_{\epsilon_{reg}}}.
\end{equation}
Physically, this field can be interpreted as the combination of the background wind and a controllable component that points along the steepest descent direction of the value function, scaled by the local admissible speed.

\subsection{Steady density transport and boundary semantics}
Given the induced velocity field \(\bm{u}(\bm{x})\), the distribution of traffic is governed by the steady-state continuity equation
\begin{equation}
\nabla\cdot(\rho\bm{u})=q(\bm{x}),\qquad \bm{x}\in\Omega_{free}\setminus\mathcal{D}.
\end{equation}
Here, \(q(\bm{x})\) is a source term determined by the scenario. In homing scenarios, \(q\) is spatially distributed to simulate uniform demand across the free space, whereas in point-to-point scenarios, it is localized to represent a specific origin region. While our default experiments assume purely advective transport, a small diffusion term \(-\kappa\Delta\rho\) can optionally be introduced for numerical regularization, in which case the total flux becomes \(\bm{J}=\rho\bm{u}-\kappa\nabla\rho\).

Correct boundary semantics are essential for interpreting the resulting density patterns. We enforce a no-flux condition on the outer domain boundaries and all obstacle surfaces, ensuring that traffic flows around rather than through these structures:
\begin{equation}
(\rho\bm{u}-\kappa\nabla\rho)\cdot\bm{n}=0,\qquad \bm{x}\in(\partial\Omega\cup\partial\Omega_{obs})\setminus\partial\mathcal{D}.
\end{equation}
Conversely, the target \(\mathcal{D}\) is treated as an open, absorbing boundary that permits inflow. In our implementation, this is realized by actively removing mass within \(\mathcal{D}\) during the density update steps. To verify the physical consistency of the simulation, we check the global mass balance between the total injected mass and the total mass absorbed at the target interface:
\begin{equation}
\int_{\Omega_{free}\setminus\mathcal{D}} q(\bm{x})\,d\bm{x}=\oint_{\partial\mathcal{D}} \rho\bm{u}\cdot\bm{n}_{\Omega_{free}\setminus\mathcal{D}}\,ds.
\end{equation}

\subsection{Assumptions and reproducibility}
Our study focuses on steady-state behavior driven by static wind fields and time-independent sources. All numerical controls related to stability, such as regularization factors, relaxation parameters, grid resolution, and solver thresholds, are managed via the \texttt{paperconfig/*.yaml} files. Furthermore, every experimental run generates a comprehensive set of artifacts, including configuration snapshots, metrics logs, and plots. This structure ensures that the results discussed in subsequent sections are fully traceable to specific, concrete run directories.

\section{Hybrid solution framework}

\subsection{Overview}
\begin{figure}[htbp]
\centering
\hspace*{-3.8cm}
\begin{tikzpicture}[
  font=\small,
  node distance=10mm and 12mm,
  box/.style={draw,rounded corners,align=center,inner sep=5pt,minimum height=10mm,fill=black!3},
  boxaccent/.style={box,fill=black!6},
  boxplain/.style={box,fill=white},
  arr/.style={-{Latex[length=2.2mm]},thick},
  arrlight/.style={-{Latex[length=2.0mm]},thick,draw=black!65}
]
\node[boxplain] (rho) {Density field\\$\rho^k$};
\node[boxaccent, right=of rho] (pinn) {PINN solve value PDE\\Eikonal/HJB for $\phi^k$\\(hard BC on $\partial\mathcal{D}$)};
\node[box, right=of pinn] (u) {Induced velocity\\$\bm{u}^k=\bm{v}_w- v_{max}(\rho^k)\,\dfrac{\nabla\phi^k}{\|\nabla\phi^k\|_{\epsilon_{reg}}}$};
\node[boxaccent, right=of u] (fvm) {Conservative FVM transport\\Solve $\nabla\cdot(\rho\bm{u}^k)=q$\\(no-flux on $\partial\Omega,\partial\Omega_{obs}$;\\absorb in $\mathcal{D}$)};

\node[boxplain, dashed, draw=black!55, fill=white, above=10mm of u, minimum width=34mm] (wind) {Wind field\\$\bm{v}_w(\bm{x})$};
\node[boxplain, dashed, draw=black!55, fill=white, above=10mm of pinn, minimum width=40mm] (barrier) {Obstacle barrier (optional)\\$\mathrm{Barrier}_{obs}(\bm{x})$};

\draw[arr] (rho) -- (pinn) node[midway,above] {$v_{max}(\rho)$};
\draw[arr] (pinn) -- (u) node[midway,above] {$\nabla\phi^k$};
\draw[arr] (u) -- (fvm);
\draw[arrlight] (fvm.south) |- ++(0,-10mm) -| (rho.south) node[pos=0.25,below,draw=black!35,rounded corners,fill=black!2,inner sep=2pt] {relaxation: $\rho^{k+1}=(1-\alpha)\rho^k+\alpha\,\tilde\rho^{k+1}$};

\draw[arrlight,dashed] (wind.south) -- (u.north);
\draw[arrlight,dashed] (barrier.south) -- (pinn.north);
\end{tikzpicture}
\caption{Overall pipeline of the hybrid framework.}
\label{fig:pipeline}
\end{figure}
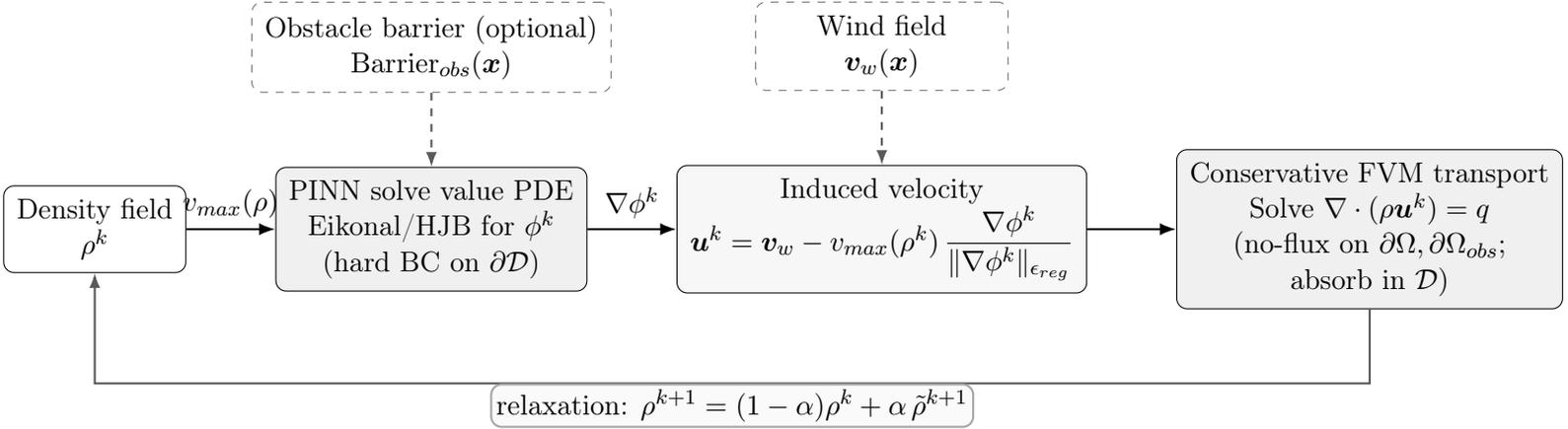

To address the coupled value-density system effectively, we employ a constraint-preserving hybrid route, referred to as Route B. This approach is driven by a pragmatic goal: to utilize a flexible solver for the anisotropic, geometry-dependent value-function PDE, while employing a conservative solver where flux bookkeeping and boundary semantics dictate the physical meaning of the density field. This strategy significantly reduces the risk of density patterns being dominated by numerical leakage or nonconservative errors. Specifically, the value-function subproblem is handled by a PINN to leverage its automatic differentiation capabilities and flexible sampling within three-dimensional space. Conversely, the density-transport subproblem is solved using a conservative Finite-Volume Method (FVM) with upwind fluxes \cite{leveque2002fvm,toro2009riemann}, which allows for explicit control over mass conservation and no-flux boundary semantics. These two distinct solvers are coupled through an outer Picard iteration with under-relaxation, creating a traceable loop \(\rho\to\phi\to\bm{u}\to\rho\) as illustrated in Fig.~\ref{fig:pipeline}.

\subsection{PINN for the value function}
During each outer iteration \(k\), we fix the density \(\rho^k\) and train a PINN to approximate the value function \(\phi^k\) by minimizing the anisotropic Eikonal residual over the free space \(\Omega_{free}\setminus\mathcal{D}\). Since the value field is primarily utilized through its gradient, we prioritize stable gradient quality by sampling interior points using geometry-aware rules, such as increasing sampling density near obstacles and the target, as specified in the configuration. To avoid the trade-offs often found between PDE residuals and boundary losses, we hard-encode the absorbing boundary condition \(\phi|_{\partial\mathcal{D}}=0\) using a nonnegative target distance function \(d_{\mathcal{D}}(\bm{x})\) combined with a Softplus output parameterization. This formulation keeps the training objective strictly focused on the PDE residual while enforcing the target semantics by construction.

When obstacles are present, we implement a specific obstacle treatment involving barrier injection and residual masking, with parameters defined in \texttt{paperconfig/*.yaml}. This strategy prevents the PINN from allocating representational capacity to invalid regions and ensures that induced gradients remain aligned with free-space navigation around obstacle boundaries. Concretely, letting \(d_{obs}(\bm{x})\) denote an obstacle signed-distance function that is positive in free space and negative inside obstacles, we define a nonnegative barrier term
\begin{equation}
\mathrm{Barrier}_{obs}(\bm{x})=C_{height}\,\mathrm{Softplus}(-k\,d_{obs}(\bm{x})).
\end{equation}
We inject this term in a manner that preserves the absorbing boundary by multiplying it with \(d_{\mathcal{D}}(\bm{x})^p\):
\begin{equation}
\hat\phi_{bc}(\bm{x})=d_{\mathcal{D}}(\bm{x})^p\,\mathrm{Softplus}(NN_\phi(\bm{x})),\qquad
\hat\phi_{total}(\bm{x})=\hat\phi_{bc}(\bm{x})+d_{\mathcal{D}}(\bm{x})^p\,\mathrm{Barrier}_{obs}(\bm{x}).
\end{equation}
Furthermore, the Eikonal residual is evaluated only on points satisfying \(d_{obs}(\bm{x})>\delta\), a configurable buffer, which enforces a residual mask consistent with the free-space semantics.

\subsection{Conservative FVM for steady density transport}
Once the velocity field \(\bm{u}^k\) is established, we solve the divergence equation \(\nabla\cdot(\rho\bm{u}^k)=q\) using a conservative finite-volume discretization on a Cartesian grid. The FVM update is formulated in flux form, ensuring that the discrete divergence corresponds exactly to the net flux across cell faces, thereby making mass balance an explicit and inspectable quantity. We employ upwind fluxes to maintain robustness in transport-dominated regimes and apply a pseudo-time relaxation method to reach a steady state. Specifically, we evolve the equation
\begin{equation}
\partial_\tau\rho+\nabla\cdot(\rho\bm{u}^k)=q
\end{equation}
until a relative change criterion is met.

Boundary semantics are imposed directly within the discrete fluxes: we enforce no-flux conditions at the outer boundary, and no-penetration at obstacles by forcing the face flux between free cells and obstacle cells to zero. Additionally, we implement absorbing target semantics by setting \(\rho=0\) inside \(\mathcal{D}\) at every pseudo-time step. Utilizing an upwind flux, setting \(\rho=0\) inside \(\mathcal{D}\) yields an ``open'' absorbing interface that permits free-to-target outflow while effectively preventing target-to-free inflow. The pseudo-time step size and iteration caps, such as CFL-style rules and \texttt{max\_iters}, are configured specifically for each scenario, and we log the number of inner iterations and residuals in \texttt{metrics.csv} to ensure traceability.

\subsection{Outer fixed-point coupling}
We alternate between solving these two subproblems and update the density using under-relaxation \(\rho^{k+1}=(1-\alpha)\rho^k+\alpha\,\mathrm{clip}(\tilde\rho^{k+1},0,\rho_{max})\), where \(\alpha\) and the stopping criteria are specified by the configuration. Clipping serves as an explicit safeguard to keep the iterate within a feasible range defined by the scenario, while relaxation mitigates the sensitivity of the coupled loop, particularly in regions with strong wind or narrow passages.

To support empirical stability and consistency checks, we report diagnostic fields in \texttt{metrics.csv}. These diagnostics include Eikonal residual statistics recorded during PINN training, density solver residuals from pseudo-time relaxation, and the \texttt{rho\_change} metric across outer iterations to quantify the magnitude of fixed-point updates. It is important to note that these diagnostics are not intended to claim theoretical convergence; rather, they provide a transparent and detailed record of solver behavior for each reported run.

\begin{algorithm}[t]
\caption{Constraint-preserving hybrid PINN--FVM solver.}
\label{alg:hybrid}
\begin{algorithmic}[1]
\STATE Initialize $\rho^0$ on $\Omega_{free}$; set $\rho=0$ in $\mathcal{D}$ and $\Omega_{obs}$.
\FOR{$k=0,1,\dots$}
\STATE Train PINN for $\phi^k$ with hard absorbing boundary on $\partial\mathcal{D}$.
\STATE Construct $\bm{u}^k$ from $\nabla\phi^k$ and $v_{max}(\rho^k)$.
\STATE Solve $\nabla\cdot(\rho\bm{u}^k)=q$ by conservative FVM to obtain $\tilde\rho^{k+1}$.
\STATE Update $\rho^{k+1}\leftarrow(1-\alpha)\rho^k+\alpha\,\mathrm{clip}(\tilde\rho^{k+1},0,\rho_{max})$.
\STATE Stop if diagnostics indicate stability (e.g., \texttt{rho\_change} below threshold).
\ENDFOR
\end{algorithmic}
\end{algorithm}

\section{Experimental design}

\subsection{Scenarios}
We evaluate the framework using two reproducible scenario families designed to induce qualitatively different flow patterns. The first is a \emph{homing} scenario, where a spatially uniform source injects mass across the free space \(\Omega_{free}\setminus\mathcal{D}\), all of which is absorbed by a single target. The second is a \emph{point-to-point} scenario, characterized by mass injection in a localized source region and subsequent absorption at the destination. To ensure full reproducibility, the geometry, source terms, and solver hyperparameters for both families are uniquely defined in external configuration files located at \texttt{paperconfig/*.yaml}. We treat these files as the authoritative specifications, mapping every reported plot and metric directly back to a concrete run directory.

\subsection{Wind and obstacle configurations}
We evaluate the robustness of our model across multiple wind types, including none, uniform, vortex, and height-dependent fields, as well as various obstacle configurations described by axis-aligned bounding boxes. In this formulation, wind is modeled as a static spatial field \(\bm{v}_w(\bm{x})\) that influences both the value-function equation and the induced motion field, while obstacles are defined as excluded regions \(\Omega_{obs}\) that enforce no-flux semantics for density transport. Although the mathematical definitions and implementation conventions adhere to the project documentation, all specific numeric parameter values are drawn directly from \texttt{paperconfig/*.yaml}. This separation allows us to discuss model semantics conceptually without cluttering the text with experiment-specific constants.

\subsection{Baselines and diagnostics}
Our primary experimental method is the hybrid framework, referred to as Route B. To provide a meaningful baseline, we compare this against an end-to-end PINN approach, Route A, which jointly learns \(\phi\) and \(\rho\) using residual and flux losses. This baseline serves as a diagnostic contrast, highlighting practical differences in how mass conservation and boundary semantics manifest in the resulting density fields. Our analysis relies on a combination of qualitative and quantitative diagnostics. Qualitative assessments include value slices, streamlines of \(\bm{u}\), and density slices that reveal structural features such as bands, bottlenecks, and splitting behavior. These are complemented by quantitative metrics extracted directly from run artifacts like \texttt{metrics.csv}, which tracks training curves and residuals, and \texttt{run\_meta.json}, which records runtime metadata. Where relevant, we further validate transport semantics using a global mass balance check that compares total injection against total absorption at the target boundary \(\partial\mathcal{D}\).

\subsection{Reproducibility}
To facilitate rigorous verification, each experimental run outputs a self-contained directory containing the full configuration, metric logs, metadata, and generated plots. Consequently, this paper reports results by referencing these specific artifacts rather than hardcoding values. This workflow ensures that it is possible to regenerate the exact plots from the provided configuration and allows readers to verify that the presented tables and figures are fully consistent with the logged diagnostics.

\section{Results}

This section presents an analysis of the computed fields illustrated in Fig.~\ref{fig:phi} through Fig.~\ref{fig:convergence}. Our examination focuses on how wind fields and obstacles reshape the value function \(\phi\), how these geometric changes translate into motion corridors, and how the resulting steady density \(\rho\) manifests as congestion bands, splitting structures, or bottlenecks. Additionally, we evaluate an end-to-end PINN baseline to illustrate the degradation that occurs when conservation laws and boundary semantics are not explicitly enforced.

\subsection{Value function under wind and obstacles}
\begin{figure}[htbp]
\centering
\includegraphics[width=\linewidth]{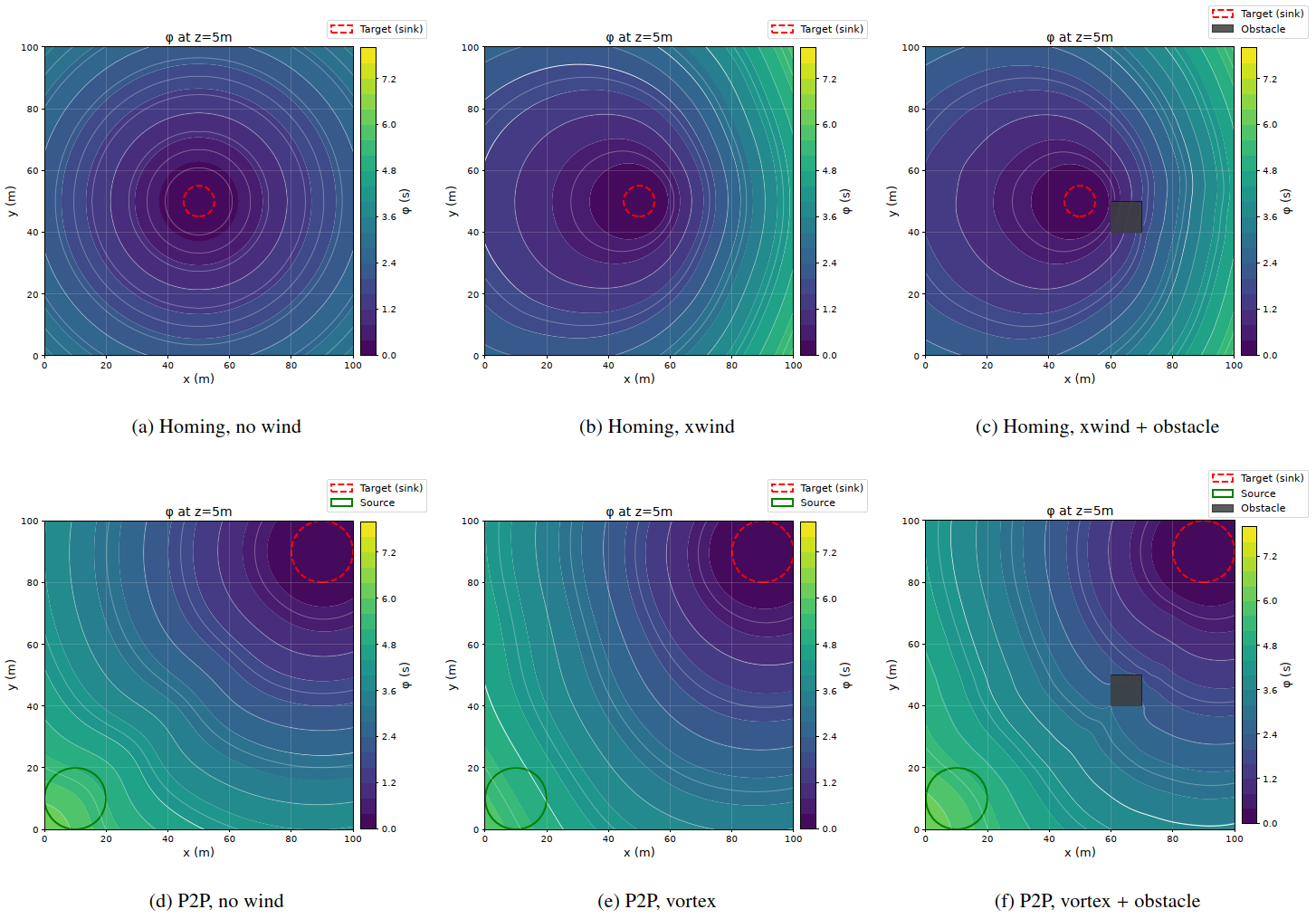}
\caption{Value function slice comparisons. Top row: Homing under baseline, crosswind, and crosswind with an obstacle. Bottom row: point-to-point (P2P) under baseline, vortex wind, and vortex wind with an obstacle. The red dashed circle marks the target (sink); the green circle marks the source (P2P panels); gray squares denote obstacles.}
\label{fig:phi}
\end{figure}

The value function \(\phi\) encodes the minimum-time reachability to the target under the influence of wind. Its level sets provide insight into how wind anisotropy alters the geometry of shortest paths and how obstacles deflect optimal travel directions. Fig.~\ref{fig:phi} compares representative slices across different tasks, wind conditions, and obstacle settings.

In the Homing task shown in the top row, the no-wind case exhibits nearly concentric level sets centered at the target. In contrast, the introduction of a crosswind stretches and shifts these contours, producing a distinct left-right asymmetry with steeper gradients on the right side. When an obstacle is placed to the right of the target, the contours are forced to bend around the block, resulting in tightened spacing near its edges that reflects a detour in minimum-time reachability. In the point-to-point (P2P) task shown in the bottom row, the baseline case forms smooth, target-centered quarter-circle contours. The vortex wind then skews these contours and breaks the radial symmetry, while the presence of an obstacle further pinches the level sets around the block, indicating a narrower effective passage toward the target. Across all settings, we observe that no spurious local minima appear away from the target or obstacle geometry, which is crucial for preventing nonphysical trapping when interpreting the induced motion.

\subsection{Induced motion patterns}
Given the value function \(\phi\) and the wind field, the induced motion \(\bm{u}\) summarizes how mass is expected to move through free space toward the target under the model semantics. Streamline visualizations allow us to identify corridor-like structures, obstacle-avoidance behaviors, and flow deflections caused by wind.

\begin{figure}[htbp]
\centering
\includegraphics[width=0.85\linewidth]{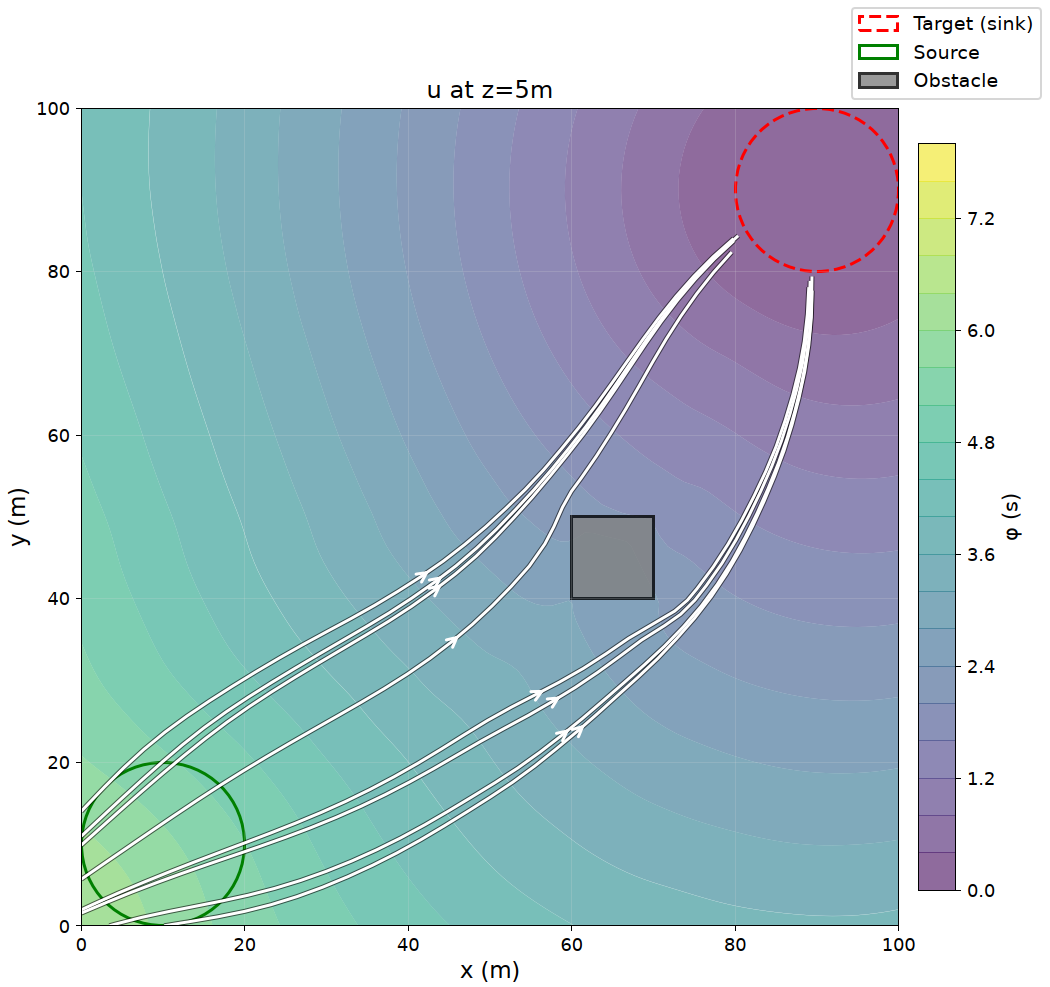}
\caption{Induced-motion streamlines (white) for P2P under vortex wind with an obstacle, overlaid on the \(\phi\) background. The target, source, and obstacle are marked.}
\label{fig:streamlines}
\end{figure}

Fig.~\ref{fig:streamlines} depicts trajectories departing from the source, curving under the influence of the vortex field, splitting to pass above and below the obstacle, and subsequently rejoining as they approach the target. The tight bundling of streamlines around the obstacle highlights narrow corridors where flow merges, effectively anticipating where density \(\rho\) will concentrate in the steady transport solution.

\subsection{Density distribution and traffic patterns}
The steady density \(\rho\) provides the most direct view of traffic patterns. Under identical geometries, different wind conditions can shift congestion hotspots and determine whether traffic splits into multiple corridors or concentrates into a single bottleneck. Because \(\rho\) is computed using the conservative transport solver in Route B, the resulting patterns are inherently consistent with the no-flux and absorbing boundary semantics.

\begin{figure}[htbp]
\centering
\includegraphics[width=0.85\linewidth]{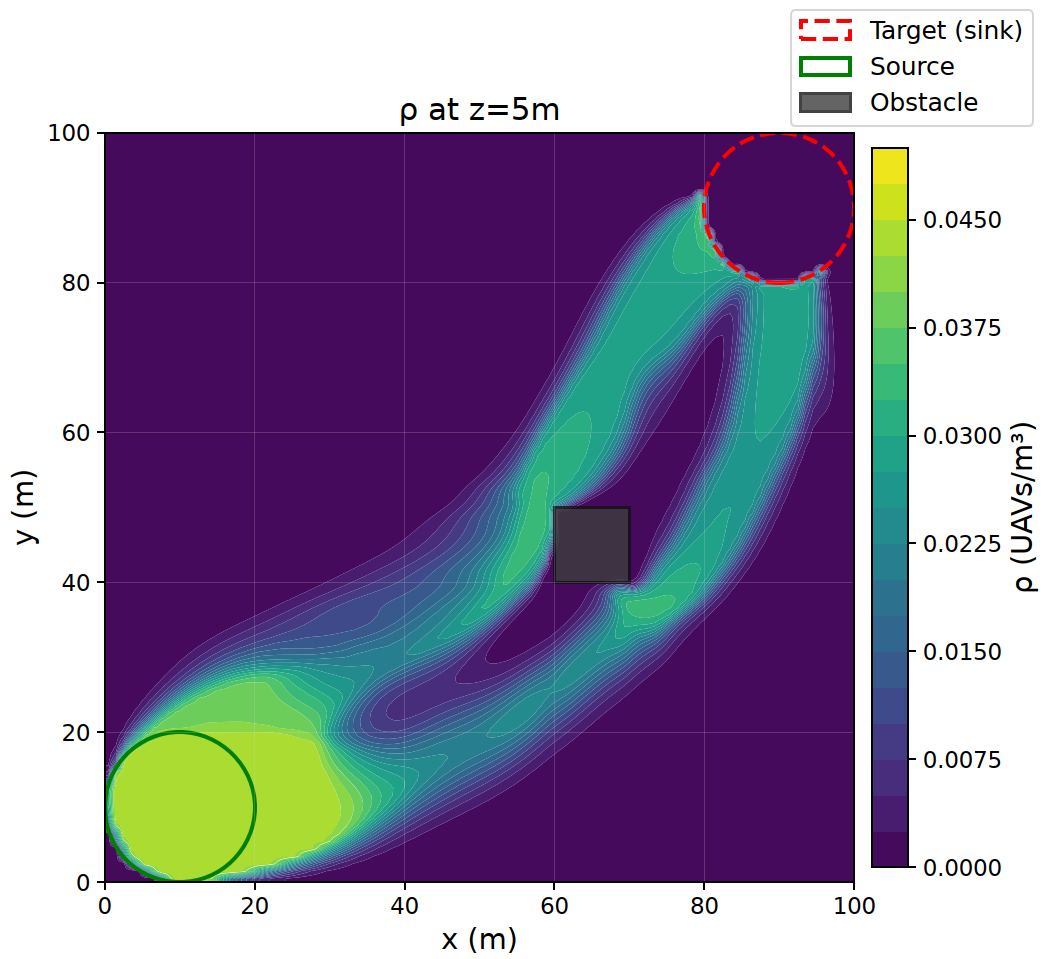}
\caption{Steady density slice (\(\rho\)) for P2P under vortex wind with an obstacle (source/target/obstacle marked).}
\label{fig:rho}
\end{figure}

Fig.~\ref{fig:rho} displays a bright density band connecting the source and target along the same curved corridor suggested by the streamlines. The obstacle forces this band to split and wrap around the block before merging downstream, producing a localized high-density neck near the effective passage. We observe that density is highest near the source region and tapers as it approaches the target, which is consistent with transport behavior under wind and geometric constraints. To ensure reliability, we cross-check that the density solver residual decreases during pseudo-time relaxation and that the global mass balance remains consistent throughout the run.

\subsection{Comparison with end-to-end PINN and FSM--FVM}
We include an end-to-end PINN baseline, referred to as Route A, to serve as a contrast case. The objective is not to claim general superiority, but rather to highlight practical differences that are critical for traffic semantics, specifically regarding conservation, boundary-flux behavior, and the stability of density patterns in transport-dominated regimes.

\begin{figure}[htbp]
\centering
\includegraphics[width=\linewidth]{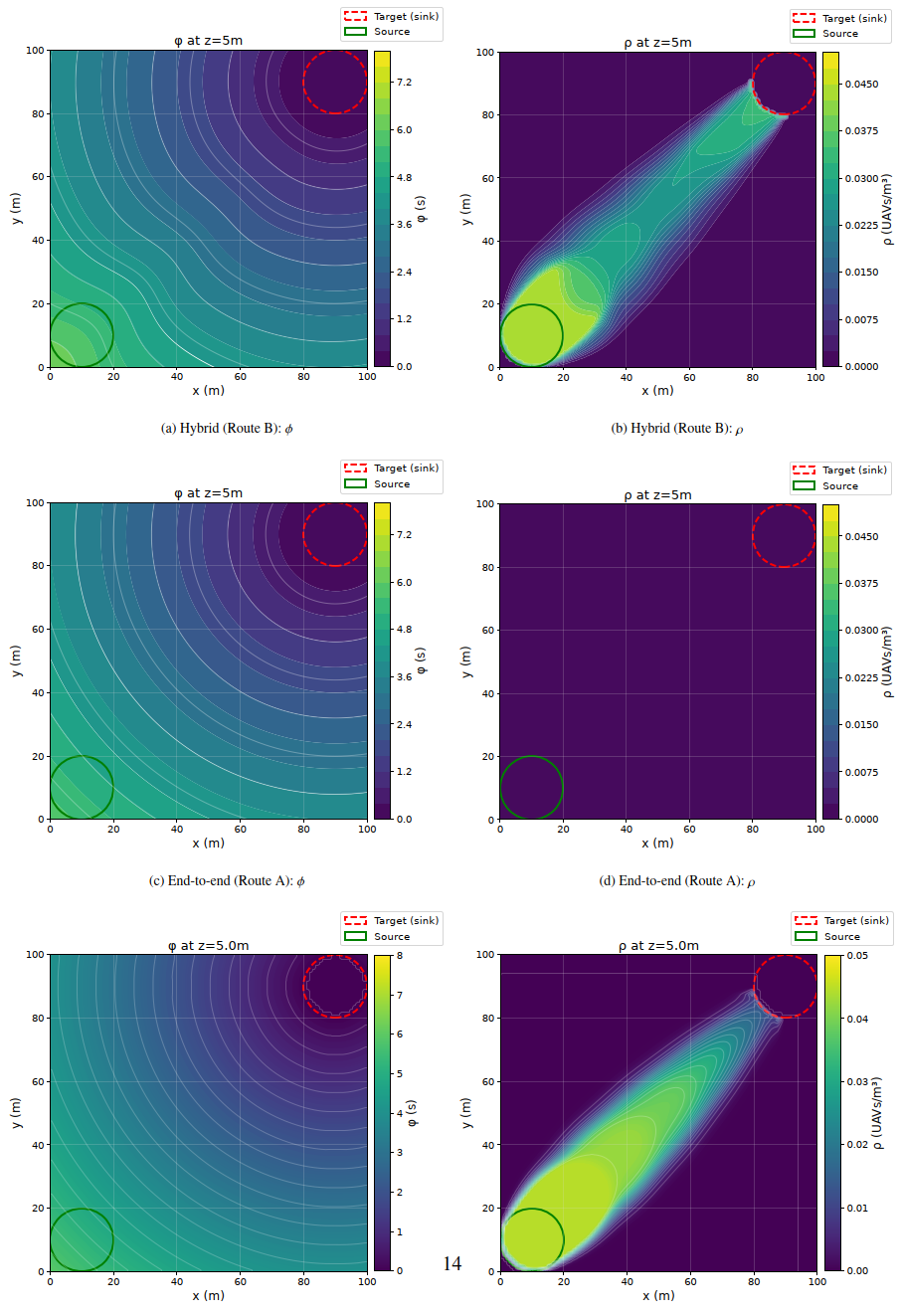}
\caption{Method comparison under the P2P scenario with no wind and no obstacles. Left column: \(\phi\); right column: \(\rho\). Rows: hybrid (Route B), end-to-end PINN (Route A), and a traditional FSM--FVM reference solver.}
\label{fig:compare}
\end{figure}

Fig.~\ref{fig:compare} compares three solvers in a benign point-to-point setting with no wind and no obstacles. The value-function slices are qualitatively consistent across methods and exhibit the expected near-radial structure around the target. However, the end-to-end baseline produces a degenerate density field that remains near zero over most of the domain, indicating that transport semantics are not captured even when the value field appears plausible. In contrast, the hybrid route, or Route B, yields a concentrated density corridor connecting the source to the target, and the resulting structure agrees qualitatively with the traditional FSM-FVM reference. This comparison underscores the practical takeaway that fitting \(\phi\) alone is insufficient for traffic interpretation; the \(\rho\) field must be validated against conservation and boundary behavior.

Using the FSM-FVM \(\phi\) as a numerical reference, we further quantify the agreement of the hybrid \(\phi\) field on a shared evaluation set of 492,984 points. The relative errors are \(L_2\) \(=8.334\times10^{-2}\), \(L_1\) \(=8.031\times10^{-2}\), and \(L_\infty\) \(=1.350\times10^{-1}\), with an \(\mathrm{RMSE}=2.479\times10^{-1}\) and range-normalized \(\mathrm{NRMSE}=4.121\times10^{-2}\). The Pearson correlation reaches 0.990, indicating that the spatial structure of \(\phi\) is highly consistent with the reference and that discrepancies are mainly localized, typically near sharp-gradient regions around boundaries or obstacles, rather than reflecting a global shape mismatch.

The training logs provide a complementary quantitative view of this failure mode. For the end-to-end run, \texttt{metrics.csv} reports a small residual-based objective at the final epoch, yet the learned \(\rho\) collapses to a near-zero field. This behavior is consistent with a trivial-solution attractor in residual minimization where \(\rho\) enters multiplicatively in the transport residual and can suppress flux-related terms. Furthermore, the loss is typically averaged over space, meaning localized source and sink constraints contribute weakly unless sampling and weighting are carefully balanced. By contrast, Route B decouples \(\phi\) learning from \(\rho\) transport and enforces conservation at the discrete flux level; in the same setting, the outer-loop summaries show a reduction of the Eikonal mean residual from approximately \(4.7\times10^{-2}\) to \(1.2\times10^{-2}\) while maintaining a non-degenerate density corridor.

\subsection{Quantitative diagnostics}
Quantitative diagnostics serve as the link between qualitative patterns and solver behavior. For Route B, \texttt{metrics.csv} records both inner-loop quantities, such as Eikonal residual statistics during PINN training, and outer-loop quantities, such as the change in density across Picard iterations. For Route A, the log typically includes the joint loss components and learning-rate schedules. We recommend reporting paired runs using the same scenario and configuration except for the route to enable a controlled comparison.

\begin{table}[t]
\centering
\caption{Wall-clock runtime comparison under multiple wind fields (no obstacles) for P2P and Homing tasks. The hybrid PINN--FVM solver is evaluated on a desktop GPU (RTX~5090). The FSM--FVM baseline is evaluated on a dual-socket CPU server (2$\times$AMD~EPYC~7T83). The results are reported as end-to-end wall-clock times under the stated platforms and should not be interpreted as a hardware-normalized algorithmic benchmark: the reported ``speedup'' reflects both platform differences and the fact that neural inference and residual evaluation are naturally GPU-parallel. ``Speedup'' is computed as $T_{\text{baseline}}/T_{\text{hybrid-GPU}}$ within each scenario; higher is better.}
\label{tab:runtime}
\small
\begin{tabular}{lrrr}
\toprule
Scenario & Hybrid (GPU) (s) & Baseline (CPU) (s) & Speedup \\
\midrule
P2P (no wind) & 288.4 & 3884.7 & 13.47 \\
P2P (xwind) & 288.3 & 2703.8 & 9.38 \\
P2P (vortex) & 312.8 & 4018.8 & 12.85 \\
\addlinespace
Homing (no wind) & 301.3 & 3078.6 & 10.22 \\
Homing (xwind) & 293.5 & 3827.5 & 13.04 \\
Homing (vortex) & 301.8 & 4289.5 & 14.21 \\
\bottomrule
\end{tabular}
\end{table}

\begin{figure}[htbp]
\centering
\includegraphics[width=\linewidth]{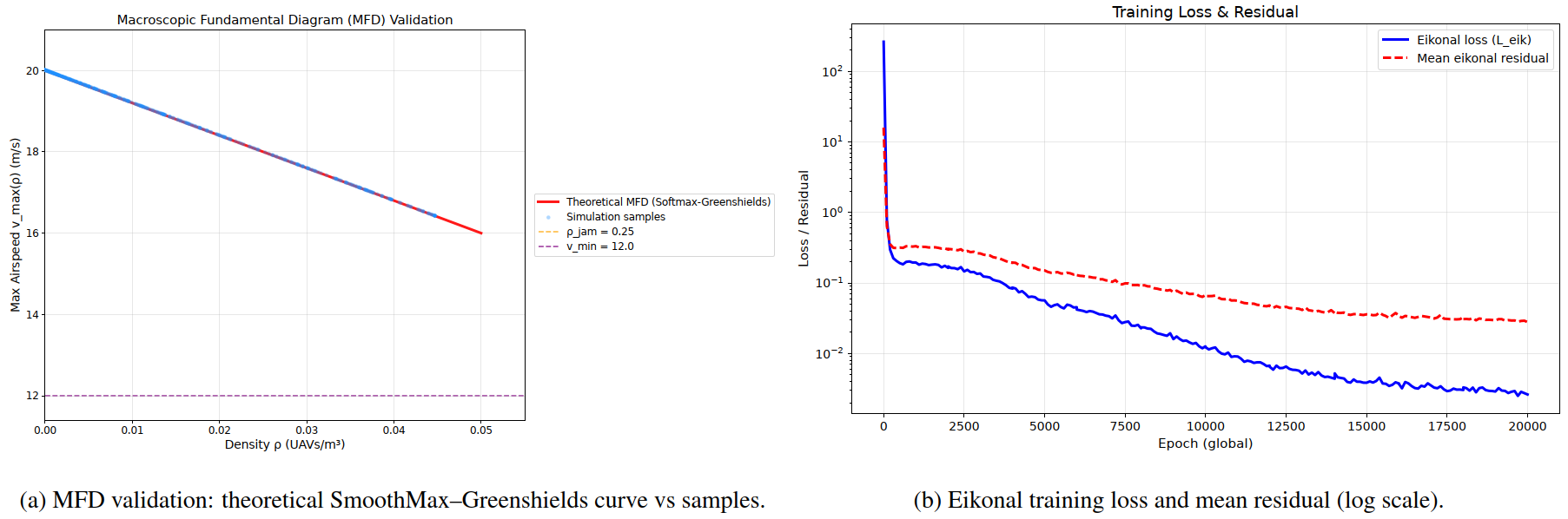}
\caption{Representative diagnostics from \texttt{metrics.csv}: MFD consistency and training convergence.}
\label{fig:convergence}
\end{figure}

As shown in Fig.~\ref{fig:convergence}, the sampled \(v_{max}(\rho)\) values align with the theoretical SmoothMax-Greenshields curve, confirming that the congestion-dependent speed mapping is respected in the run. Furthermore, the Eikonal loss and mean residual drop rapidly and then plateau at small values, providing a run-level consistency check rather than a theoretical proof of convergence.

\section{Discussion and implications}

\subsection{Why the hybrid design is effective }
The effectiveness of the hybrid split is motivated by a mechanism-level view of the coupled system rather than a purely theoretical claim. The value-function partial differential equation benefits significantly from the flexibility of the physics-informed neural network in three-dimensional space, particularly when handling complex coefficients and geometry-aware sampling. Conversely, the transport equation benefits from a conservative discretization that directly controls mass conservation and enforces no-flux boundary semantics. This distinction is critical because steady density patterns rely heavily on flux consistency; if mass is not conserved or if obstacle boundaries leak, qualitative structures such as bottlenecks may be mere numerical artifacts. The outer Picard iteration with relaxation effectively couples these two subproblems while allowing each inner solver to remain specialized. Furthermore, the recorded diagnostics, including Eikonal residual statistics, density solver residuals, and the outer-loop \texttt{rho\_change}, provide a comprehensive run-level record. These metrics indicate whether the solver has entered a consistent regime and help identify runs that are dominated by numerical transients.

\subsection{Interpreting density patterns as traffic phenomena}
Steady density structures, such as bands, bottlenecks, and merging or splitting patterns, provide traffic-relevant interpretations for airspace corridor organization and congestion hotspots under environmental disturbances \cite{aarts2023capacity}. In our setting, these structures can be interpreted as the macroscopic outcome of a population following locally induced motion influenced by wind and obstacles, with \(v_{max}(\rho)\) serving as the coupling mechanism between congestion and achievable speed. From a traffic-management perspective, these fields support several operational questions, such as identifying where corridors emerge naturally under a given wind field and how they shift when wind patterns change. Additionally, the framework helps determine which passages become bottlenecks in the presence of obstacles and assesses their sensitivity to different demand patterns, such as homing versus point-to-point sources. It also allows analysts to distinguish whether congestion hotspots are caused by geometric constraints like narrow passages or by wind-induced anisotropy such as directional drift. Ultimately, the goal is not to replace detailed safety analysis, but to provide a fast and interpretable diagnostic layer that facilitates scenario comparison.

\subsection{Limitations and failure modes}
The current study is intentionally scoped to examine steady-state behavior under static wind fields, and as a result, it does not capture transient demand surges, time-varying wind, or vehicle-level dynamics. On the numerical side, the coupled loop can be sensitive to relaxation parameters, grid resolution, and sampling density near narrow passages or regions with strong wind. Furthermore, the obstacle treatments in the physics-informed neural network are engineered and configuration-dependent, which may require tuning when obstacles are extremely tight or when the wind field creates sharp anisotropy. Finally, the macroscopic model abstracts away specific safety separation standards, conflict-resolution rules, and heterogeneous vehicle capabilities. While these aspects are important for real-world deployment, they motivate future extensions that can integrate additional constraints while retaining the reproducible, diagnostics-driven workflow established in this paper.

\section{Conclusion}

We have presented a reproducible, constraint-preserving hybrid PINN--FVM framework designed for macroscopic UAV traffic flow in three-dimensional domains characterized by wind and obstacles. By decoupling the geometry-dependent value function from the density transport problem and linking them through an outer fixed-point iteration, we demonstrated that combining a flexible PINN representation for anisotropic value-function PDEs with a conservative transport solver successfully preserves the intended semantics of boundary conditions and mass balance. This preservation is critical when interpreting density patterns as genuine traffic phenomena rather than numerical artifacts, as evidenced by our empirical results showing how wind fields and obstacles reshape value structures and organize steady density. Building on this foundation, future research will focus on non-stationary extensions that account for time-varying wind and dynamic demand, alongside the integration of realistic vehicle constraints and scalable acceleration strategies to support broader comparative evaluations.

\section*{Disclosures} 
The authors declare no conflicts of interest.

\section*{Acknowledgements}
This work is supported by the developing Project of Science and Technology of Jilin Province (20250102032JC).

\section*{Code availability}
The code for this work is available as open source at \url{https://github.com/ToughClimb/uav-mfg-hybrid} under the Apache License 2.0.

\end{document}